\documentclass{ws-procs975x65}

\begin{document}
\begin{flushright}
YITP-SB-13-3
\end{flushright}
\vspace{-0.2cm}

\title{On the developments of Sklyanin's quantum separation of variables for integrable quantum field theories.}
\author{G. NICCOLI}
\address{C.N. YANG INSTITUTE FOR THEORETICAL PHYSICS, STONY BROOK UNIVERSITY, STONY BROOK, NY 11794-3840, USA, E-mail: niccoli@max2.physics.sunysb.edu}
\begin{abstract}We present a microscopic approach in the framework of Sklyanin's quantum separation of variables (SOV) for the exact solution of 1+1-dimensional quantum field theories by integrable lattice regularizations. Sklyanin's SOV is the natural quantum analogue of the classical method of separation of variables and it allows a more symmetric description of classical and quantum integrability  w.r.t. traditional Bethe ansatz methods. Moreover, it has the advantage to be applicable to a more general class of models for which its implementation gives a characterization of the spectrum complete by construction. Our aim is to introduce a method in this framework which allows at once to derive the spectrum (eigenvalues and eigenvectors) and the dynamics (time dependent correlation functions) of integrable quantum field theories (IQFTs). This approach is presented for a paradigmatic example of relativistic IQFT, the sine-Gordon model.\end{abstract}
\vspace{-0.1cm}
\bodymatter
\section{Introduction}
The solution of quantum field theories by the complete characterization of their spectrum (eigenvalues and eigenstates) and dynamics (time dependent correlation functions) is one of the fundamental issues in mathematical physics as it should lead to exact (non-perturbative) results in several areas of physics where these models play a central role. The 1+1-dimensional
case\cite{ICMP12Heise28} is the most
natural framework where to try to solve exactly this problem thanks to the
powerful tools of quantum integrability\cite{ICMP12Yang67,ICMP12FadST79} . Despite significant progresses obtained in the last forty years for some lattice models (like the Heisenberg spin chains), the full solution of more general integrable quantum field theories (IQFTs) is still a fundamental open problem.  Our
main aim is to define a microscopic approach for the exact solution of 1+1-dimensional quantum field theories by integrable lattice regularizations in the framework of the quantum inverse scattering
method (QISM). In this framework, the quantum integrable structure:
\begin{equation}
\mathsf{T}(\lambda )\in \mathsf{End}(\mathcal{H}):[\mathsf{T}(\lambda ),%
\mathsf{T}(\mu )]=0\text{ \ }\forall \lambda ,\mu \in \mathbb{C},\text{ \ }%
H\in \mathsf{T}(\lambda )
\end{equation}\vspace{-0.5cm}
of a quantum model of Hamiltonian $H\in $\textsf{End}$(\mathcal{H})$ on the
quantum (Hilbert) space $\mathcal{H}$ is defined by the one-parameter family
of \textit{transfer matrices}:
\begin{equation}
\mathsf{T}(\lambda )\equiv \text{tr}_{\mathbb{C}^{2}}\mathsf{M}_{0}(\lambda
),\text{ \ \ \ \ }\mathsf{M}_{0}(\lambda )\equiv \left( 
\begin{array}{cc}
\mathsf{A}(\lambda ) & \mathsf{B}(\lambda ) \\ 
\mathsf{C}(\lambda ) & \mathsf{D}(\lambda )%
\end{array}%
\right) _{0}\in \mathsf{End}(\mathcal{H}\otimes \mathbb{C}^{2}).  \label{Tdef}
\end{equation}%
Here, we are restricting ourselves to a \textit{monodromy matrix }$\mathsf{M}(\lambda )\,\in \mathsf{End}(\mathcal{H}\otimes \mathbb{C}^{n})$ with $n=2$; this is a solution of
the so-called Yang-Baxter equation in \textsf{End}$(\mathcal{H}\otimes 
\mathbb{C}^{2}\otimes \mathbb{C}^{2})$:
\begin{equation}
R_{0,0^{\prime }}(\lambda /\mu )\,(\mathsf{M}_{0^{{}}}(\lambda )\otimes
1_{0^{\prime }})\,(1_{0^{{}}}\otimes \mathsf{M}_{0^{\prime }}(\mu
))\,=\,(1_{0^{{}}}\otimes \mathsf{M}_{0^{\prime }}(\mu ))\,(\mathsf{M}%
_{0^{{}}}(\lambda )\otimes 1_{0^{\prime }})R_{0,0^{\prime }}(\lambda /\mu ),
\label{YBA}
\end{equation}
and $R_{0,0^{\prime }}(\lambda )\in $\textsf{End}$(\mathbb{C}^{2}\otimes 
\mathbb{C}^{2})$ is a solution of the Yang-Baxter equation in \textsf{End}$(%
\mathbb{C}^{2}\otimes \mathbb{C}^{2}\otimes \mathbb{C}^{2})$. The elements of $%
\mathsf{M}(\lambda )$ are the generators of a representation
in $\mathcal{H}$ of the Yang-Baxter algebra and, for invertible $R$-matrix, the commutation
relations (\ref{YBA}) imply the mutual commutativity of the one-parameter
family of transfer matrices $\mathsf{T}(\lambda )$.
\subsection{Local fields identification problem in S-matrix formulation}
It is worth recalling that some classes of massive integrable quantum field
theories (IQFTs) in infinite volume can be defined avoiding a microscopic
lattice regularization. Indeed, they admit an \textit{on-shell}\cite{ICMP12Zam77} exact and complete characterization by the exact S-matrices which fixes the
asymptotic particle dynamics. The main difficulty here is that any information needs to be extracted from the particle dynamics. In particular, a direct connection between
local fields and form factors (their matrix elements on asymptotic particle states) is absent and the form factors are characterized axiomatically as solutions of a set of functional equations\cite{ICMP12KarW78} completely fixed by the exact S-matrix. Different methods have addressed this longstanding
problem and the description of massive IQFTs as (superrenormalizable)
perturbations of conformal field theories by relevant local fields\cite{ICMP12Zam89} has characterized one important research line. The
consequent hypothesis of isomorphism of the local field content between
massive theories and the corresponding ultraviolet conformal ones has been
verified for some fundamental IQFTs both for the chiral\cite{ICMP12Kub95} and the non-chiral local fields\cite{ICMP12DN05-1} by form factor analysis. These are important results on the global
structure of the local operator spaces of the massive IQFTs but they do not
really lead to the identification of particular local fields. It is worth
recalling that in\,\,\,\cite{ICMP12BabBS96} a criterion based on the quasi-classical
characterization of the local fields has been introduce to define the correspondence between local fields and form factors. It has been fully
described in the special cases of the restricted sine-Gordon model at the
reflectionless points for chiral fields and verified on the basis of
counting arguments\footnote{It is worth mentioning that the new fermionic structures described in\,\,\,\cite{ICMP12BJMST07-02} , appearing from the lattice regularization given in terms of the XXZ
spin-1/2 quantum chain, have been used recently to investigate the structure
of form factors of the sine-Gordon model in the infinite volume limit.
Remarkably the authors of\,\,\,\cite{ICMP12BJMST11} were able to reproduce the results of the papers\,\,\,\cite{ICMP12BabBS96} from this different approach.}. From the above discussion it is then clear that in the
S-matrix formulation the main open problem remains the absence of a direct 
reconstruction of the quantum local fields.
\subsection{\label{schema}Integrable microscopic approach in SOV framework}
One of our main motivations to develop an integrable microscopic approach to
quantum field theories is to introduce an exact setup where to overcome the
identification problem in the S-matrix formulation. Indeed, in the QISM
framework, we can use the so-called solution of the quantum inverse
scattering problem, a central achievements in the Lyon group method\cite{ICMP12KitMT99} , which applies to a large class of lattice
integrable quantum models and allows to write explicitly the local operators
in terms of the global Yang-Baxter generators. Such a result
plays a key role in the derivation of multiple integral representations of
correlation functions as it is at the basis of the algebraic computations of
the local operator actions on the transfer matrix eigenstates. The Lyon
group method has been develop by using the algebraic Bethe
ansatz (ABA) as central tool for the spectrum characterization. However, ABA does not work
for important integrable quantum models on the contrary of the Sklyanin's
quantum separation of variables (SOV)\cite{ICMP12Skl85}\,. This
beautiful method is quite general and powerful to describe the spectrum of
these models; it leads to both the eigenvalues and the eigenstates of the
transfer matrix with a spectrum construction (which under simple conditions)
has as built-in feature its completeness. Moreover, in the SOV framework,
for the so far analyzed quantum models\,\,\,\cite{ICMP12NicT10,ICMP12NicT10+,ICMP12GMN12-SG} it was an easy task to prove
their complete quantum integrability; i.e. the simplicity of the transfer
matrix spectrum. Our aim is to develop a method based on the Sklyanin's SOV
which exactly characterize the spectrum and the dynamics (correlation
functions) of IQFTs according to the following general schema:
\begin{description}
\item[\textsf{A)}] Solution of the spectral problem, for the lattice and the
continuum theories: \textsf{A1)} Solution of the spectral problem for the
integrable lattice regularizations; i.e. SOV construction of transfer matrix
eigenstates and eigenvalues. \textsf{A2)} Reformulation of the spectrum in
terms of nonlinear integral equations (of thermodynamical Bethe ansatz type)
and definition of finite volume quantum field theories by continuum limit. 
\textsf{A3)} Derivation of S-matrix description of the spectrum in the IR limit, infinite volume. \textsf{A4)} Derivation of the renormalization group
fixed point conformal spectrum in the UV limit.\smallskip 

\item[\textsf{B)}] Exact formulae for the correlation functions: \textsf{B1)}
Reconstruction of the local operators in terms of the Sklyanin's quantum separate
variables. \textsf{B2)} Determinant form for the scalar product of the class of {\it separate states}, which contains also the transfer matrix eigenstates. \textsf{B3) }Matrix elements of local
operators on transfer matrix eigenstates. \textsf{B4)} Thermodynamical limit
and derivation of multiple integral formulae for correlation functions.
\end{description}

\subsection{On the Sklyanin's quantum separation of variables}

Following Sklyanin\cite{ICMP12Skl85} , we would like first to present a possible
definition of quantum separate variables for an integrable quantum model.
Let $Y_{n},P_{n}\in $ \textsf{End}$(\mathcal{H})$ be $\mathsf{N}$ couples of
canonical conjugate operators $[Y_{n},Y_{m}]=[P_{n},P_{m}]=0,$ $%
[Y_{n},P_{m}]=\delta _{n,m}/2\pi i\ $and let us assume that $\{Y_{1},...,Y_{%
\mathsf{N}}\}$ are simultaneously diagonalizable operators with simple spectrum on $%
\mathcal{H}$. Then, we can present the following:

\begin{description}
\item[\textbf{Definition}] \textit{The set of operators }$\{Y_{1},...,Y_{%
\mathsf{N}}\}$\textit{\ define the quantum separate variables for the
spectral problem of the one parameter family of conserved charges }$\mathsf{T%
}(\lambda )$\textit{\ if and only if there exist (under an appropriate
definition of the operator order) }$\mathsf{N}$\textit{\ quantum separate
relations of the type:}%
\begin{equation}
F_{n}(Y_{n},P_{n},\mathsf{T}(Y_{n}))=0\text{ \ \ \ }\forall n\in \{1,..,%
\mathsf{N}\}.  \label{Separ-relations}
\end{equation}
\end{description}
These are quantum analogues of the classical separate relations in the
Hamilton-Jacobi's approach and are here used to solve the spectral
problem of $\mathsf{T}(\lambda )$. Thanks to (\ref{Separ-relations}), in the eigenbasis of $\mathcal{H}$ formed out of $\{Y_{1},...,Y_{\mathsf{N}}\}$ simultaneous eigenstates $|y_{1},....,y_{N}\rangle $ with $y_{n}$ being the $Y_{n}$-eigenvalues, the generic $\mathsf{T}(\lambda )$-eigenstate $|t\rangle $ with eigenvalue $t(\lambda )$ is characterized by the following separate equations:
\begin{equation}
F_{n}(y_{n},\frac{i}{2\pi }\frac{\partial }{\partial y_{n}},t(y_{n}))\Psi
_{t}(y_{1},....,y_{N})=0,\ \ \text{where }\Psi
_{t}(y_{1},....,y_{N})=\langle y_{1},....,y_{N}|t\rangle\,,  \label{FrbtSOVBax1}
\end{equation}%
$\forall
n\in \{1,..,\mathsf{N}\}$. Then it is natural searching
for wave function solutions of factorized form\footnote{Note that  an
independent proof of the completeness of the above factorized ansatz is required.} $\Psi
_{t}(y_{1},....,y_{N})=\prod_{n=1}^{\mathsf{N}}Q_{t}^{(n)}(y_{n})$, where $%
Q_{t}^{(n)}(y_{n})$ is a solution of the equations (\ref{FrbtSOVBax1}) for
the fixed $n\in \{1,...,\mathsf{N}\}$. One of the fundamental contributions of Sklyanin\cite{ICMP12Skl85} has been to define a procedure to determine the quantum
separate variables in the framework of QISM for the transfer matrix spectral
problem and the explicit form of the corresponding quantum separate
relations. In the class of integrable quantum models defined by a \textit{%
monodromy matrix} $\mathsf{M}(\lambda )$ of the form (\ref{Tdef}) this
procedure reads:

\begin{description}
\item[\textbf{Sklyanin's procedure to SOV}] \textit{If the generator} $\mathsf{B}(\lambda )$ \textit{of the Yang-Baxter algebra defines a one parameter family of simultaneously diagonalizable commuting operators with
simple spectrum then the operator zeros }$\{Y_{1},...,Y_{\mathsf{N}}\}$%
\textit{\ of }$\mathsf{B}(\lambda )$ \textit{define the quantum separate
variables for the transfer matrix spectral problem. Moreover, the
corresponding }$\mathsf{N}$ \textit{separate equations are Baxter like
second order difference equations computed in the spectrum of each quantum
separate variable.}
\end{description}
\section{The sine-Gordon model}
\subsection{Classical model}
The classical sine-Gordon model can be characterized by the Hamiltonian
density H$_{SG}\equiv \left( \partial _{x}\phi \right) ^{2}+\Pi ^{2}+8\pi
\mu \cos 2\beta \phi $, where the field $\phi (x,t)$ is defined for $%
(x,t)\in \lbrack 0,R]\otimes \ \mathbb{R}$ with periodic boundary conditions $%
\phi (x+R,t)=\phi (x,t)$. The dynamics of the model in the Hamiltonian
formalism is defined in terms of $\phi (x,t),$ $\Pi (x,t)$ with $\{\Pi
(x,t),\phi (y,t)\}=2\pi \delta (x-y)$. The classical integrability of the
sine-Gordon model is assured thanks to the representation of the equation of
motion by a zero-curvature condition $[\partial _{t}-V(x,t;\lambda ),\partial
_{x}-U(x,t;\lambda )]=0$. Here, we have defined $U=$k$_{1}\sigma _{1}\cos
\beta \phi +$k$_{2}\sigma _{2}\sin \beta \phi -$k$_{3}\sigma _{3}\Pi ,$ $V=-$%
k$_{2}\sigma _{1}\cos \beta \phi -$k$_{1}\sigma _{2}\sin \beta \phi -$k$%
_{3}\sigma _{3}\partial _{x}\phi $ and k$_{1}=i\beta \left( \pi \mu \right)
^{1/2}(\lambda -\lambda ^{-1}),$ k$_{2}=i\beta \left( \pi \mu \right)
^{1/2}(\lambda +\lambda ^{-1}),$ k$_{3}\equiv i\beta /2$ and the $\sigma _{a}
$ are the standard Pauli matrices.
\subsection{Quantum lattice regularization}
In order to regularize the ultraviolet divergences that arise in the
quantization of the model a lattice discretization can be introduced. The
field variables are discretized according to the standard recipe $\phi
_{n}\equiv \phi (n\Delta )$\ and $\Pi _{n}\equiv \Delta \Pi (n\Delta )$,
where $\Delta =R/N$ is the lattice spacing. Then, the canonical quantization
is defined by imposing that $\phi _{n}$\ and $\Pi _{n}$ are self-adjoint
operators satisfying the commutation relations $[\phi _{n},\Pi _{m}]=2\pi
i\delta _{n,m}$. The quantum lattice regularization of the sine-Gordon model here used goes back to\,\,\,\cite{ICMP12FadST79} and
it is characterized by the following Lax operator:%
\begin{equation}
\mathsf{L}_{n}(\lambda )=\kappa _{n}\left( 
\begin{array}{cc}
\mathsf{u}_{n}(q^{-1/2}\mathsf{v}_{n}\kappa _{n}+q^{1/2}\mathsf{v}%
_{n}^{-1}\kappa _{n}^{-1}) & (\lambda _{n}\mathsf{v}_{n}-(\mathsf{v}%
_{n}\lambda _{n})^{-1})/i \\ 
(\lambda _{n}/\mathsf{v}_{n}-\mathsf{v}_{n}/\lambda _{n})/i & \mathsf{u}%
_{n}^{-1}(q^{1/2}\mathsf{v}_{n}\kappa _{n}^{-1}+q^{-1/2}\mathsf{v}%
_{n}^{-1}\kappa _{n})%
\end{array}%
\right)   \label{Lax}
\end{equation}%
where $\lambda _{n}\equiv \lambda /\xi _{n}$ for any $n\in \{1,...,\mathsf{N}%
\}$ and $\xi _{n}$ and $\kappa _{n}$ are the parameters of the model. Here,
the basic operators are the unitary operators $\mathsf{v}_{n}\equiv
e^{-i\beta \phi _{n}}$ and $\mathsf{u}_{n}\equiv e^{i\beta \Pi _{n}/2}$
which generate $\mathsf{N}$ independent local Weyl algebras $\mathsf{u}_{n}%
\mathsf{v}_{m}=q^{\delta _{nm}}\mathsf{v}_{m}\mathsf{u}_{n}\,$, with\ \
parameter $q\equiv e^{-i\pi \beta ^{2}}$, thanks to $[\phi _{n},\Pi
_{m}]=2\pi i\delta _{n,m}$. Then the monodromy matrix that characterize
the lattice sine-Gordon model is $\mathsf{M}(\lambda )\equiv \mathsf{L}_{%
\mathsf{N}}(\lambda )\cdots \mathsf{L}_{1}(\lambda )$ and it satisfies the
Yang-Baxter equation w.r.t. the standard 6-vertex $R$-matrix.

\subsection{Cyclic representations}

Here, we restrict our attention to the case in which $q$ is a $p$-root of
unity, e.g. $\beta ^{2}\,=p^{\prime }/p$ with $p$ odd and $p^{\prime }$ even
coprime. This implies that the powers $p$ of the generators $\mathsf{u}_{n}$
and $\mathsf{v}_{n}$ are central elements of each local Weyl algebra. In
this case, we can associate a $p$-dimensional linear space $\mathcal{H}_{n}$
to any site $n$ of the lattice and we can define on it the following cyclic
representation of the Weyl algebra:
\begin{equation}
\mathsf{v}_{n}|k_{n}\rangle =q^{k_{n}}|k_{n}\rangle ,\text{ \ }\mathsf{u}%
_{n}|k_{n}\rangle =|k_{n}-1\rangle ,\text{ \ }|k_{n}+p\rangle =|k_{n}\rangle
,\text{ }\forall k_{n}\in \{0,...,p-1\}.  \label{v-eigenbasis}
\end{equation}
Then, the quantum space associated to the lattice sine-Gordon model is the $%
p^{\mathsf{N}}$-dimensional Hilbert space $\mathcal{H}\equiv \otimes _{n=1}^{%
\mathsf{N}}\mathcal{H}_{n}$. In these representations the following
definition $\mathcal{O}(\lambda ^{p})\,\equiv \,\prod_{k=1}^{p}\mathsf{O}%
(q^{k}\lambda )\,$ of average of a one-parameter family of commuting
operators $\mathsf{O}(\lambda )$ plays a very important role; indeed it holds:
\begin{proposition}[\cite{ICMP12NicT10}]
The average of the monodromy matrix elements are central in the
Yang-Baxter algebra and are characterized by:%
\begin{align}
\mathcal{B}(\Lambda )& =\mathcal{C}(\Lambda )=\left( \text{F}(-\Lambda )-%
\text{F}(\Lambda )\right) /2,\text{ \ \ \ }\mathcal{A}(\Lambda )=\mathcal{D}%
(\Lambda )=\left( \text{F}(-\Lambda )+\text{F}(\Lambda )\right) /2\,,
\label{Averages} \\
\text{F}(\Lambda )& \equiv \prod_{r=1}^{\mathsf{N}}\left( \kappa _{r}\xi
_{r}/i\right) ^{p}(1+(-1)^{p^{\prime }/2}i^{p}\left( \kappa _{r}/\xi
_{r}\right) ^{p}\Lambda )(1+(-1)^{p^{\prime }/2}i^{p}\Lambda /\left( \kappa
_{r}\xi _{r}\right) ^{p})/\Lambda .
\end{align}
\end{proposition}
\section{Transfer matrix spectral problem solution in SOV}
In\,\,\,\cite{ICMP12NicT10} the spectrum of the lattice sine-Gordon model has been
completely characterized by SOV, we briefly summarize the results which
mainly define the step \textsf{A1} in the schema defined in subsection \ref{schema}. In order to simplify the notations we will describe here and in
the following only the case odd $\mathsf{N}$. 

\subsection{Implementation of Sklyanin's SOV in sine-Gordon model}

In\,\,\,\cite{ICMP12NicT10} the recursive construction of the spectrum (eigenvalue and
eigenstates) of the\textit{\ }$\mathsf{B}(\lambda )$ Yang-Baxter generator
has been implemented and the main results are here reproduced:

\begin{proposition}[\cite{ICMP12NicT10}]
\textit{The operator zeros }$\{Y_{1},...,Y_{\mathsf{N}}\}$\textit{\ of }$%
\mathsf{B}(\lambda )$ are simultaneously diagonalizable and with simple
spectrum \textit{almost for all the values of the parameters of the Lax
operators. Then, they define proper Sklyanin's quantum separate variables
for the sine-Gordon spectral problem. Moreover, due to the centrality in the
Yang-Baxter algebra of }$\{Y_{1}^{p},...,Y_{\mathsf{N}}^{p}\}$\textit{,
their spectrum is completely characterized by the identities }$%
Y_{n}^{p}=Z_{n}\in \mathbb{C}$ \ $\ \forall n\in \{1,...,\mathsf{N}\}$. The $%
Z_{n}$ are the zeros of the known Laurent polynomial $\mathcal{B}(\Lambda )$ defined in 
(\ref{Averages}) while the decomposition of the identity in the SOV basis is
fixed by the following covectors-vector actions\cite{ICMP12GMN12-SG} :%
\begin{equation}
\langle y_{1}^{(k_{1})},...,y_{\mathsf{N}}^{(k_{\mathsf{N}%
})}|y_{1}^{(h_{1})},...,y_{\mathsf{N}}^{(h_{\mathsf{N}})}\rangle
=\prod_{n=1}^{\mathsf{N}}\delta _{k_{n},h_{n}}\prod_{1\leq b<a\leq \mathsf{N}%
}(y_{a}^{(h_{a})}/y_{b}^{(h_{b})}-y_{b}^{(h_{b})}/y_{a}^{(h_{a})})^{-1},
\label{M_jj}
\end{equation}%
$\forall\, k_{n}\,,h_{n}\in \{0,...,p-1\}$, where $y_{n}^{(k_{n})}\equiv
y_{n}^{(0)}q^{k_{n}}$ and $y_{n}^{(0)}$ is a fixed $p$-root of $Z_{n}$.
\end{proposition}

\subsection{Complete transfer matrix spectrum characterization}

Let us define the Laurent polynomials in $\lambda\,,$ $d(\lambda )=q^{\mathsf{N}}a(-\lambda q)$ and $a(\lambda )\equiv \prod_{r=1}^{%
\mathsf{N}}(\kappa _{r}\xi _{r}/i\lambda )(1+iq^{-\frac{1}{2}}\lambda \kappa
_{r}/\xi _{r})(1+iq^{-\frac{1}{2}}\lambda /\kappa _{r}\xi _{r})$, then it holds:

\begin{proposition}[\cite{ICMP12NicT10}]
The spectrum of the transfer matrix $\mathsf{T}(\lambda )$ is simple and the
set $\Sigma _{\mathsf{T}}$ of the eigenvalues functions coincides with the
set of $t(\lambda )$ solutions of the Baxter equation:%
\begin{equation}
t(\lambda )Q_{t}(\lambda )=a(\lambda )Q_{t}(\lambda q^{-1})+d(\lambda
)Q_{t}(\lambda q),\ \ \forall \lambda \in \mathbb{C},  \label{EQ-Baxter-R}
\end{equation}%
in the class of functions $\lambda ^{(\mathsf{N}-1)}t(\lambda )\in \mathbb{R}%
[\lambda ^{2}]_{\mathsf{N}-1}\,,Q_{t}(\lambda )\in \mathbb{R}[\lambda ]_{%
\mathsf{N}(p-1)}$, where $\mathbb{R}[\lambda ]_{M}$ is the linear space of 
\textit{real} polynomials of degree $\leq M$ in $\lambda $. Moreover, the
unique (up to normalization) eigenstate $|t\rangle $ corresponding to $t(\lambda )\in \Sigma _{\mathsf{T}}$ is characterized by\cite{ICMP12NicT10,ICMP12GMN12-SG}\,:
\begin{equation}
|t\rangle =\sum_{h_{1},...,h_{\mathsf{N}}=1}^{p}\prod_{a=1}^{\mathsf{N}%
}Q_{t}(y_{a}^{(h_{a})})\prod_{1\leq b<a\leq \mathsf{N}%
}(y_{a}^{(h_{a})}/y_{b}^{(h_{b})}-y_{b}^{(h_{b})}/y_{a}^{(h_{a})})|y_{1}^{(h_{1})},...,y_{%
\mathsf{N}}^{(h_{\mathsf{N}})}\rangle .  \label{Qeigenstate-odd}
\end{equation}
\end{proposition}
\vspace{-0.15cm}
\section{Matrix elements of local operators}
In\,\,\,\cite{ICMP12GMN12-SG} the steps \textsf{B1,} \textsf{B2} and \textsf{B3} of the
schema defined in subsection \ref{schema} have been derived, we briefly
summarize the final results on matrix elements:
\vspace{-0.1cm}
\begin{proposition}[\cite{ICMP12GMN12-SG}]
There exists a basis $\mathbb{B}_{\mathcal{H}}$ in \textsf{End}$(\mathcal{H})$
such that for any $\mathsf{O}\in \mathbb{B}_{\mathcal{H}}$ the matrix
elements on the transfer matrix eigenstates read: 
\begin{equation}\label{Gen-MELO}
\langle t^{\prime }|\mathsf{O}|t\rangle =\det_{\mathsf{N}}||\Phi
_{a,b}^{\left( \mathsf{O},t^{\prime },t\right) }||\text{, \ }\Phi _{a,b}^{\left( 
\mathsf{O},t^{\prime },t\right) }\equiv (y_{a}^{(0)})^{2b-1}\sum_{c=1}^{p}F_{\mathsf{O}%
,b}(y_{a}^{(c)})Q_{t}(y_{a}^{(c)})Q_{t^{\prime }}(-y_{a}^{(c)})q^{(2b-1)c},
\end{equation}%
where the coefficients $F_{\mathsf{O},b}(y_{a}^{(c)})$ characterize the
operator $\mathsf{O}$ and are computed by using the solution of the
quantum inverse problem. Let us show two examples:\\
a) If $\mathsf{O}$ is the identity operator, it holds $F_{\mathsf{u}%
_{1},b}(y_{a}^{(c)})=1$ \ for any\ $a,\,b\in \{1,...,\mathsf{N}\}$.\\
b) If $\mathsf{O}\equiv \mathsf{u}_{1}$ is the
Weyl algebra local generator in site $1$, it holds:%
\begin{eqnarray}
F_{\mathsf{u}_{1},b}(y_{a}^{(c)}) &=&y_{a}^{(c)}\qquad \forall b\in \{1,...,\mathsf{N}-1\},\,\,\, \forall a\in \{1,...,\mathsf{N}\}\\
F_{\mathsf{u}_{1},\mathsf{N}}(y_{a}^{(c)}) &=&\frac{\left(
y_{a}^{(0)}\right) ^{2(\mathsf{N}-1)}q^{1/2}\xi _{1}q^{(c+1)(\mathsf{N}-1)}Q_{t}(y_{a}^{(c)})Q_{t^{\prime
}}(-y_{a}^{(c+1)})}{\prod_{n=2}^{\mathsf{N}%
}\kappa _{n}/i(q(\xi _{1}\kappa
_{1})^{2}+(y_{a}^{(c+1)})^{2})Q_{t^{\prime }}(-y_{a}^{(c)})}a(\eta
_{a}^{(c+1)}).
\end{eqnarray}
\end{proposition}
\vspace{-0.3cm}
\section{Conclusion and outlook}
Let us complete this contribution evidencing the fundamental feature of
universality which emerges in the characterization of both the spectrum and
the dynamics by our approach in SOV. Indeed, this appears clearly by the
analysis of several others fundamental integrable quantum models associated
to more general cyclic representations, to highest weight
representations of 6-vertex and dynamical 6-vertex as well as to spin-1/2 representations of 8-vertex
Yang-Baxter algebra and of general 6-vertex reflection algebra.
Indeed, the results derived in\,\,\,\cite{ICMP12NicT10,ICMP12NicT10+,ICMP12GMN12-SG} show that a part from model dependent
features, like the nature of the spectrum of the quantum separate variables,
the coefficients in the Baxter equation and the SOV-reconstruction of local
operators, the spectrum and dynamics admit the same type of representations.
In particular, the form factors are expressible as determinants of matrices
with elements the \textquotedblleft convolutions\textquotedblright\ over the
spectrum of the separate variables of Baxter equation solutions plus
contributions coming from the local operators; i.e. (\ref{Gen-MELO}) seems universal. The next natural step is to complete  for the sine-Gordon model the described integrable
microscopic program. Indeed, the main point
to complete is \textsf{B4} which defines the form factors in the infinite
volume limit and allows for the comparison with those in the
S-matrix formulation in this way solving the local field identification
problem. Finally, the most intriguing projects are related to the
generalization of this analysis to more advanced quantum models like those
associated to higher rank quantum spin chains of fundamental interest in
gauge theories and also to models like Hubbard model which is a
celebrated model in condensed matter theory as it describes both the charge
and the spin degrees of freedom.
\vspace{-0.4cm}
{\small{\small\section*{Acknowledgments}
\vspace{-0.1cm}The author would like to gratefully acknowledge the fundamental contributions given by J. Teschner in\,\,\,\cite{ICMP12NicT10} and N. Grosjean and M.-J. Maillet in\,\,\,\cite{ICMP12GMN12-SG} to the subjects here described. I gratefully acknowledge the YITP Institute of Stony Brook for the opportunity to develop my research programs under support of National Science Foundation grants PHY-0969739.}}
\vspace{-0.8cm}
\bibliographystyle{ws-procs975x65}
{}
\end{document}